# Event-based Scheimpflug LiDAR for Ultra-Fast Laser-Scanned Rangefinding

NATHAN MERAZ[1,†], ALISHA WHITEHEAD[1,2,†], SUET YING CHAN[1,3,†], RONAN TANEJA[1,4], GABRIELLA MAYREND[1,5], AND JOSEPH L. GREENE[1,*]

[1]*Georgia Tech Research Institute, 250 14th Street, NW, Atlanta, GA 30318, USA*

[2]*University of Arizona, Wyant College of Optical Sciences, 1630 E. University Blvd, Tucson, AZ, 85721, USA*

[3]*Boston University, Department of Electrical and Computer Engineering, 8 St. Mary St, Boston, MA, 02215, USA*

[4]*Georgia Institute of Technology, George W. Woodruff School of Mechanical Engineering, 801 Ferst Dr NW, Atlanta, Georgia, 30332, USA*

[5]*Embry-Riddle Aeronautical University, Department of Mechanical Engineering, 600 S. Clyde Morris Blvd, Daytona Beach Florida, 32114, USA*

**Joseph.Greene@gtri.gatech.edu*
*[†]Authors contributed equally*

**Abstract:** Frame-based ranging systems are constrained by frame rate and provide no intrinsic mechanism for background rejection, limiting utility in high-throughput or cluttered environments. We present eSCHORTY, a Scheimpflug LiDAR integrating an event-based sensor with a modulated continuous-wave line laser to enable dense 3D point clouds, generated from over one million megaevents per second. We demonstrate that laser modulation provides a trade-off between event-space feature detection and localization, and that logarithmic event encoding suppresses the reflectance-induced centroid artifact demonstrated in intensity-based ranging. Reconstructions of natural scenes confirm spatially coherent depth recovery, with the Scheimpflug geometry supporting adaptation from millimeter- to kilometer-scale applications.

## 1. Introduction

Optical rangefinding, or the non-contact measurement of surfaces at range with optical precision, is a foundational capability to motivate 3D-aware systems across fields such as remote sensing [1], industrial metrology [2], topographic survey [3,4], autonomous navigation [5], and infrastructure mapping [6–8]. Active optical systems incorporate a controlled illumination source to broadly encode depth directly into the measurement [9] through four methods: multi-view, temporal, interferometric, and wavefront-encoding.

Multi-view approaches, including optical triangulation [11] and stereo vision [12], recover depth geometrically from multiple viewpoints but sacrifice compactness and necessitate measurement fusion. Time-of-flight (ToF) LiDAR [13] estimates depth from round-trip pulse travel time but demands time-gating hardware, high SWaP-C, and introduces multipath artifacts. Interferometric systems [14,15] yield high-resolution surface topographies but are constrained by range-resolution trade-offs, decorrelation, and large multi-path configurations. Wavefront-encoding methods [16–23], including point spread function engineering and optical multiplexing, encode axial information through controlled defocus through pupil modification but require custom optics and computationally intensive reconstruction. While these methods have found broad adoption, the enumerated trade-offs highlight the need for a simple methodology that does not sacrifice SWaP-C nor signal conditioning to achieve reliable 3D reconstruction at operationally relevant ranges.

Scheimpflug imaging [24] offers a compact, single-view (e.g., monocular) alternative that encodes depth geometrically without requiring time-gating or custom optics. By tilting the image plane relative to the lens, the Scheimpflug condition maps a tilted object plane to a sharp focus arising from a continuous plane in range. Pairing this geometry with a modulated continuous-wave (CW) line laser constrains the illumination to a single depth-encoded slice to enable pixel-wise range recovery along the illuminated profile. This principle enables prior Scheimpflug LiDAR [25], profilometry [26,27], bathymetry [28], and atmospheric sensing [29] systems. However, prior reliance on frame-based cameras imposes a hard throughput ceiling set by frame rate and pixel readout and remains sensitive to unstructured background arising off the illuminated plane.

Event-based sensors (EBS) [30] overcome both limitations by asynchronously reporting only the pixels that undergo a change in signal contrast exceeding a user-defined threshold with microsecond-scale resolution [31,32]. This architecture naturally suppresses static background and low-contrast defocus, remains sensitive to an expanded dynamic range through a logarithmic intensity encoding, and supports event rates up to a million events per second without the readout overhead of full-frame reporting. These properties are directly complementary to an active Scheimpflug platform where the modulated laser produces a periodic contrast signal that the EBS encodes asynchronously, while the contrast threshold acts as an intrinsic background filter, retaining returns from the in-focus illuminated plane. The result is a depth-encoding architecture that is compact, high-throughput, and selective.

Here we present eSCHORTY, an event-based Scheimpflug LiDAR that integrates these principles into a low-cost, off-the-shelf platform capable of dense 3D point cloud generation at event-camera throughput rates. We characterize the system's ranging geometry and calibration procedure, demonstrate a tunable trade-off between event-space feature detection and localization as a function of laser modulation, and quantify the influence of surface reflectance on centroid estimation. Point cloud reconstruction is accelerated using the SENPI event-based processing engine [33]. We validate eSCHORTY across controlled calibration targets and natural scenes and discuss how its ranging geometry scales readily from millimeter to kilometer distances with minimal reconfiguration [29].

## 2. Methods

### *2.1. Event-Based Scheimpflug LiDAR*

When the object and image planes of an optical system are not parallel, a sharp focus cannot be maintained across the full field under standard imaging conditions. The Scheimpflug principle resolves this scenario by requiring that the object plane, lens plane, and image plane share a common line of intersection, known as the Scheimpflug intersection point, where focus is maintained across the tilted geometry [24,34]. This principle maps a tilted object plane to a tilted image plane, allowing object features at different depths to be imaged simultaneously, as shown in **Fig. 1**. The tilt angles (θ, θ') set the operational range and resolution trade-off, with larger tilts extending the working distance of the system at the cost of compressing a greater depth range into each pixel bin.

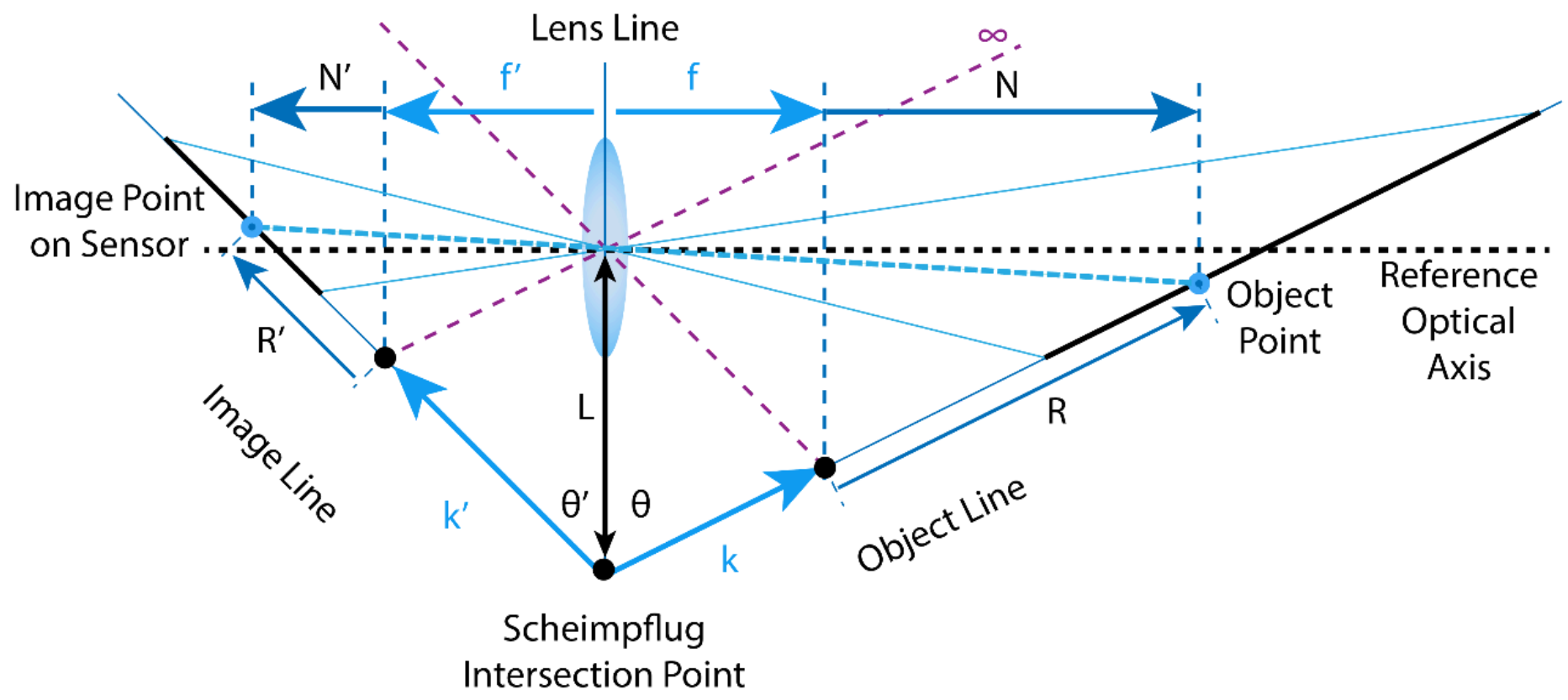

Fig. 1: Newtonian imaging equation variables (f, f’, N, N’) with the Scheimpflug geometry projections (k, k’, R, R’) used for the rangefinding equation. L is the baseline between the lens center and the Scheimpflug intersection point.

Following the ranging framework of Meraz et al. [29], adapted here for active EBS illumination, the pixel-to-range mapping relates the illuminated object plane range position R to the sensor plane coordinate R' through

$$R' = \frac{kk'}{R} = mk' \qquad (1)$$

where m is the range-dependent transverse magnification, and $k = f/\sin\theta$ and $k' = f'/\sin\theta'$ are the projections of the front and back focal lengths along the tilted object and image planes, respectively. This expression is the Scheimpflug analog of the Newtonian imaging condition, with tilted-plane projections replacing the standard on-axis distances. As a result, each pixel on the sensor maps to a unique object-plane range, except as $R' \rightarrow 0$ where the mapping diverges to infinity. A full derivation is provided in the Supplementary Material.

Integrating an EBS into this geometry adds two capabilities that frame-based cameras cannot provide. First, the EBS reports only pixels where the log-intensity change exceeds a user-defined contrast threshold, suppressing returns from out-of-focus and static background regions without the need for downstream filtering. Second, the asynchronous, timestamped event stream captures the modulated laser return at microsecond-scale resolution by removing any full-frame readout constraints. Together, these properties allow the contrast threshold to act as a depth-selective gate, which may preferentially select events generated at the in-focus laser line, where the modulated illumination strength may be estimated and exhibit sharper contrast than the defocus background. After calibration, each event pixel index can be assigned to a range bin directly, and we may leverage the spatiotemporal continuity of the scanned laser profile to suppress uncorrelated noise. Scanning the laser line orthogonally across the scene completes the 3D volume in a push-broom configuration [35], analogous to a 2D LiDAR sweep.

The Prophesee EVK4 sensor used here supports a minimum inter-event interval of approximately 1/7 µs per pixel, as reported by the manufacturer, to give a theoretical maximum throughput approaching mega-events per second (MEPS) across an illuminated slice. In practice, event generation is shaped by on-chip filtering operations, including antiflicker, frequency passbands, and the refractory period, which set the modulation response. Events are aggregated into 2D time surfaces and 3D spatiotemporal volumes using the SENPI processing engine [33] to extract the laser return from background and recover per-scan-point depth statistics. A general overview of the processing workflow is shown in **Fig. 2.**

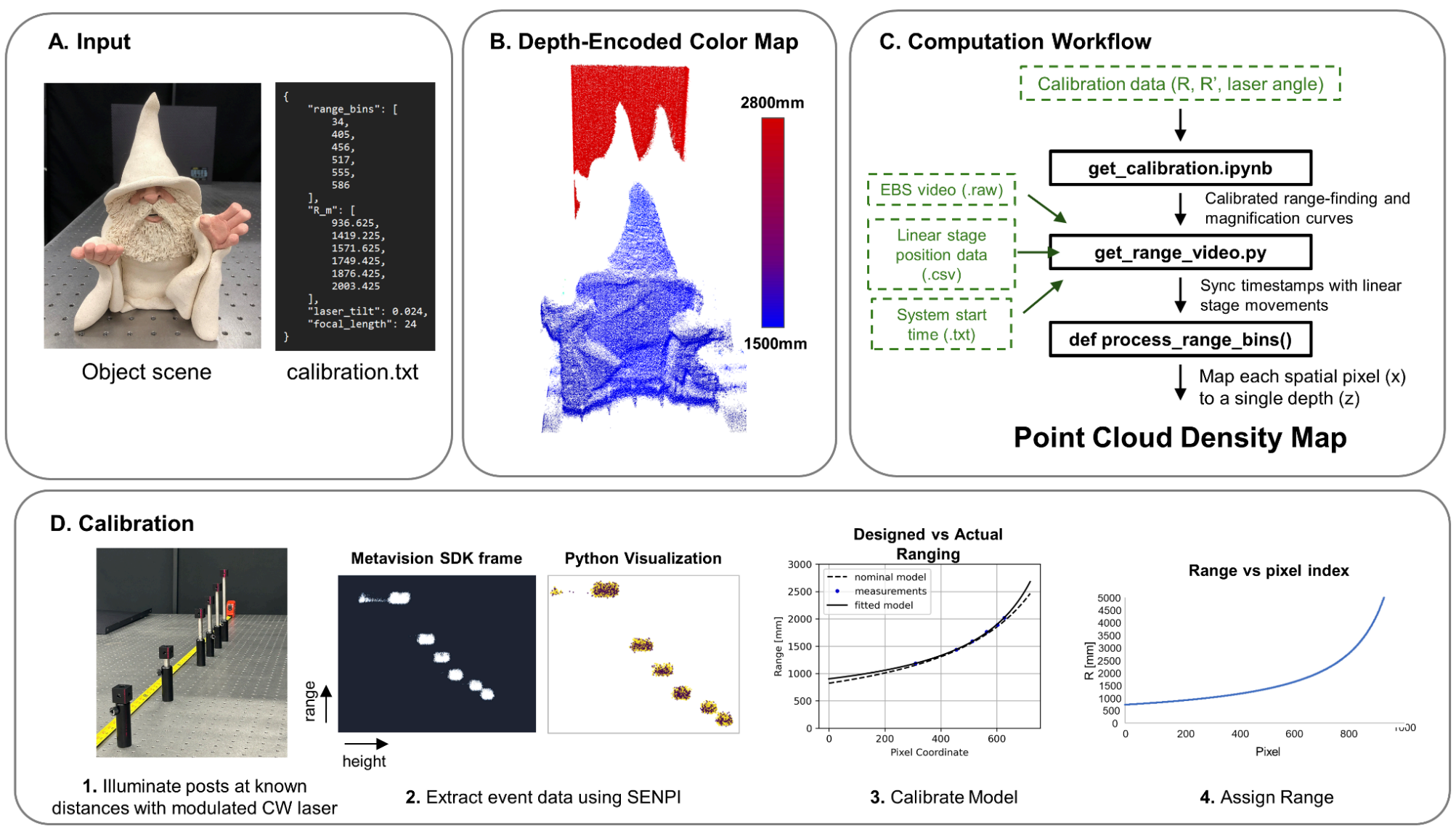

Fig. 2: Overview of point cloud generation using eSCHORTY. (A) Inputs: 3D scene to be scanned and the calibration.txt file specific to the build, (B) output: 3D point cloud, (C) block diagram of code workflow for generating the .pcd file, (D) calibration procedure that involves 1) setting up objects at known distances, 2) processing the event tensors from single-scan of the calibration scene (note: the vertical axis of the depth dimension), and 3) outputting the calibrated pixel-to-range equation.

## *2.2. Calibration*

The ranging model of Eq. (1) contains two system parameters: the projected focal lengths k and k', whose realized values depend on the physical tilt angles and the lens properties of the assembled system. Assembly tolerances, 3D-printed mount variations, and optical misalignment introduce deviations from the nominal design that must be characterized to enable accurate ranging. The calibration procedure described here estimates these parameters by mapping known object distances to their corresponding pixel coordinates in the aggregated event frame (e.g., time surface) using SENPI.

To account for practical measurement offsets, we extend Eq. (1) to include constant bias terms:

$$R' + b' = \frac{kk'}{R+b} \tag{2}$$

where b corrects for the offset between the mechanical datum and the front focal plane, and b' indexes the sensor's zero-pixel coordinate to the nearest measurable range. While k could in principle absorb correlated errors from sensor offset, projected plane angular shift, and on-axis magnification simultaneously, k, k', and b are strongly coupled, with multiple parameter combinations producing identical R-to-R' mappings and making the unconstrained optimization ill-posed. By fixing b and k to their nominal design values and applying a least-squares fit solely to the remaining free parameters, we mitigate this redundancy. This choice is validated by directly locating the front focal plane, registering the object plane tilt angle, and verifying the on-axis magnification for each configuration, confirming that the nominal values of k and b are physically realized and valid within the assembled system.

To calibrate eSCHORTY, optical posts are arranged at known positions along the object plane using the illumination laser as a reference, with posts set to varying heights to distribute the calibration targets across the angular field of view. The assembled camera is placed with the Scheimpflug intersection axis aligned to the object plane and rotated about this axis until all targets maintain sharp focus through depth. An aggregated event frame is then generated by accumulating events over the full scan duration to produce a depth-informed Scheimpflug image of the scene in which the horizontal axis encodes angular height and the vertical axis encodes range, as shown in **Fig. 3**.

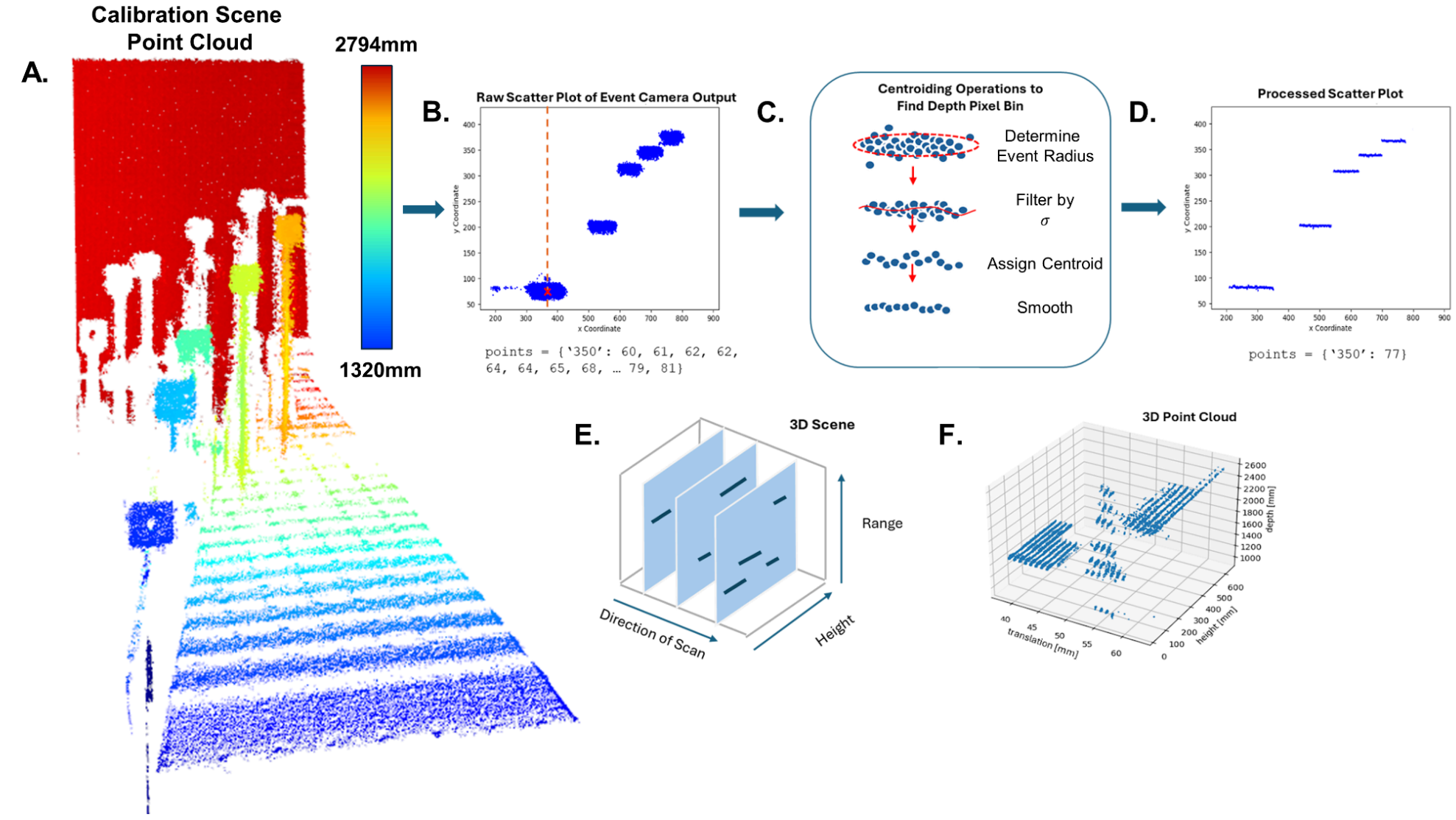

Fig. 3: (A) Event-space point cloud of calibration scene using spatially separated irises or variable height as known targets. (B) Raw scatter plot of event camera data aggregated over the observation time from the calibration scene where x is pixel row coordinate and y is height. (C) Processing pipeline uses event-based statistics to select centroid for each x. (D) Processed scatter plot where each x key corresponds to a single y value. (E) 2D frames are collected for each scan position to yield (height, range). (F) Example point cloud of calibration scene

From this frame, the centroided pixel coordinate R' is extracted for each post and fit to **Eq. (2)** to recover the calibrated pixel-to-range mapping. The depth-dependent magnification m = R'/k' is then computed and compared against the known physical sizes of the posts at their measured ranges to verify k. If the derived magnification deviates from the expected curve, k is updated accordingly until the magnification matches the experimental geometry.

During prototyping, three eSCHORTY configurations were assembled using different 3D-printed lens mounts to mate the F/2 24 mm focal length lens to the Prophesee EVK4 sensor. Config. 3 was adopted as the final design based on mechanical stability. The calibration procedure was applied to each configuration independently, and the results are reported in **Table 1** which uses pixel units of 4.86 microns for k' and b'. While k remains fixed at its design value across all configurations, k' and b' vary between builds to reflect real differences in sensor positioning introduced by the printed mounts, which the calibration procedure successfully recovers. This observation demonstrates the procedure's utility as a rapid quality check during iterative hardware development.

Table 1: eSCHORTY Model Parameters

| Name | k [mm] | k' [pix] | b [mm] | b' [pix] | theta [deg] | theta' [deg] | kk' [mm-pix] | R_near [mm] | R_far [mm] |
|---|---|---|---|---|---|---|---|---|---|
| **Nominal Design** | 24.06 | -36510.29 | 0.00 | -1084.00 | 85.87 | -7.77 | -878437 | 810.45 | 2413.54 |
| **Config. 1** | 24.06 | -32409.56 | 0.00 | -1086.04 | 85.87 | -8.76 | -779774 | 718.07 | 2124.69 |
| **Config. 2** | 24.06 | -37413.82 | 0.00 | -1075.30 | 85.87 | -7.58 | -900176 | 837.23 | 2526.73 |
| **Config. 3** | 24.06 | -37261.86 | 0.00 | -1142.74 | 85.87 | -7.62 | -896502 | 784.61 | 2115.93 |

The resulting pixel-to-range maps are shown in **Fig. 4A**, where calibration measurements fall cleanly along their respective model curves. Residual ranging errors, shown in **Fig. 4B**, remain within the geometric depth-of-field limit for all configurations, confirming that the calibration model captures the dominant sources of ranging uncertainty. For a Scheimpflug system, a model for the projected depth-of-field uncertainty along the range axis is:

$$\Delta R_{defocus} \simeq \pm \frac{p \cos(\theta') F/\#}{m^2 \sin(\theta)} \qquad (3)$$

where p is the pixel width (4.86 μm for the EVK4 sensor).

The linear relationship between R' and magnification m, shown in **Fig. 4C**, provides a complementary diagnostic, wherein the slope of each line equals 1/k', the y-intercept gives the magnification at the nearest range, and the x-intercept identifies the pixel corresponding to R = ∞. Applying the calibrated model to the iris targets in **Fig. 3** confirms agreement between the measured angular extents and the known physical iris diameters to provide an independent cross-check on the recovered value of k.

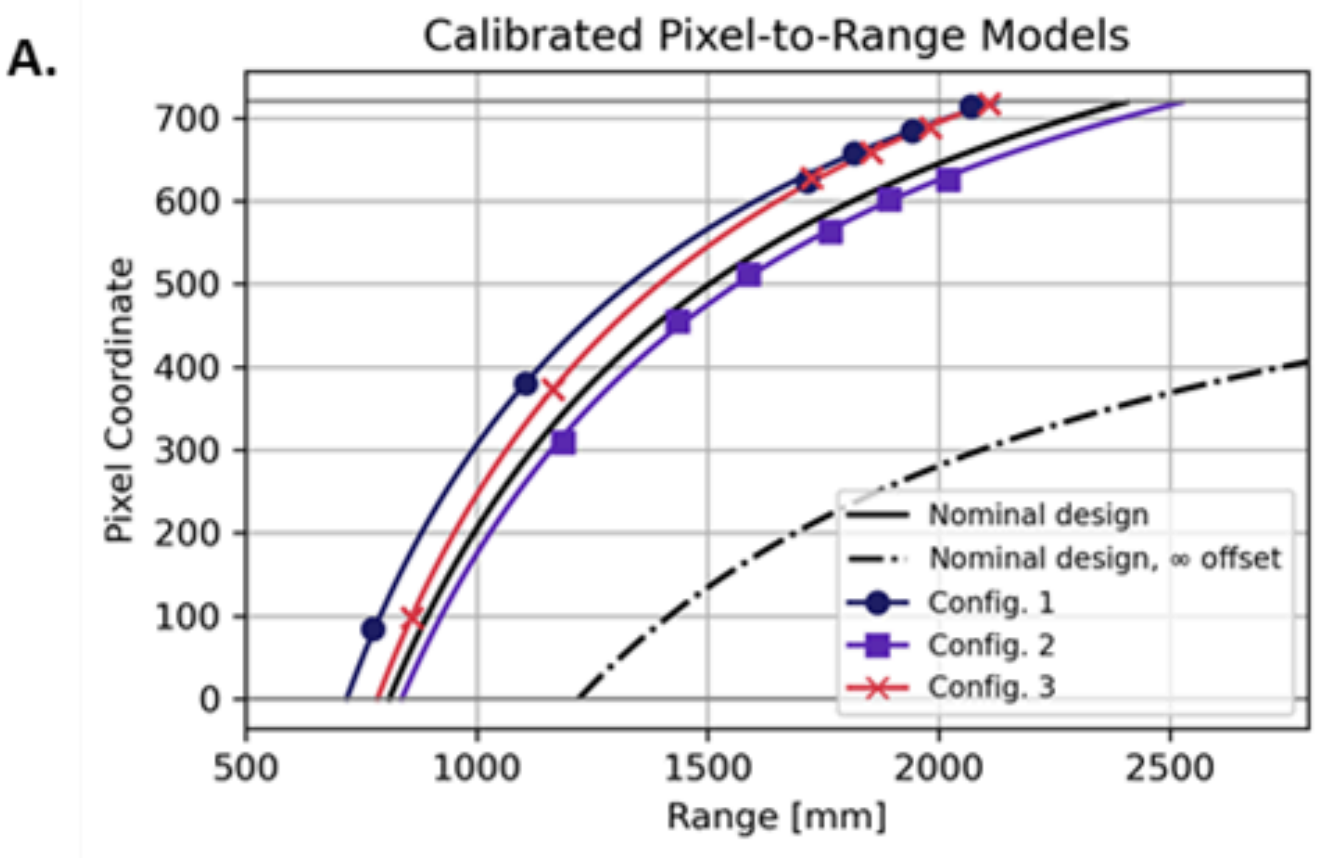

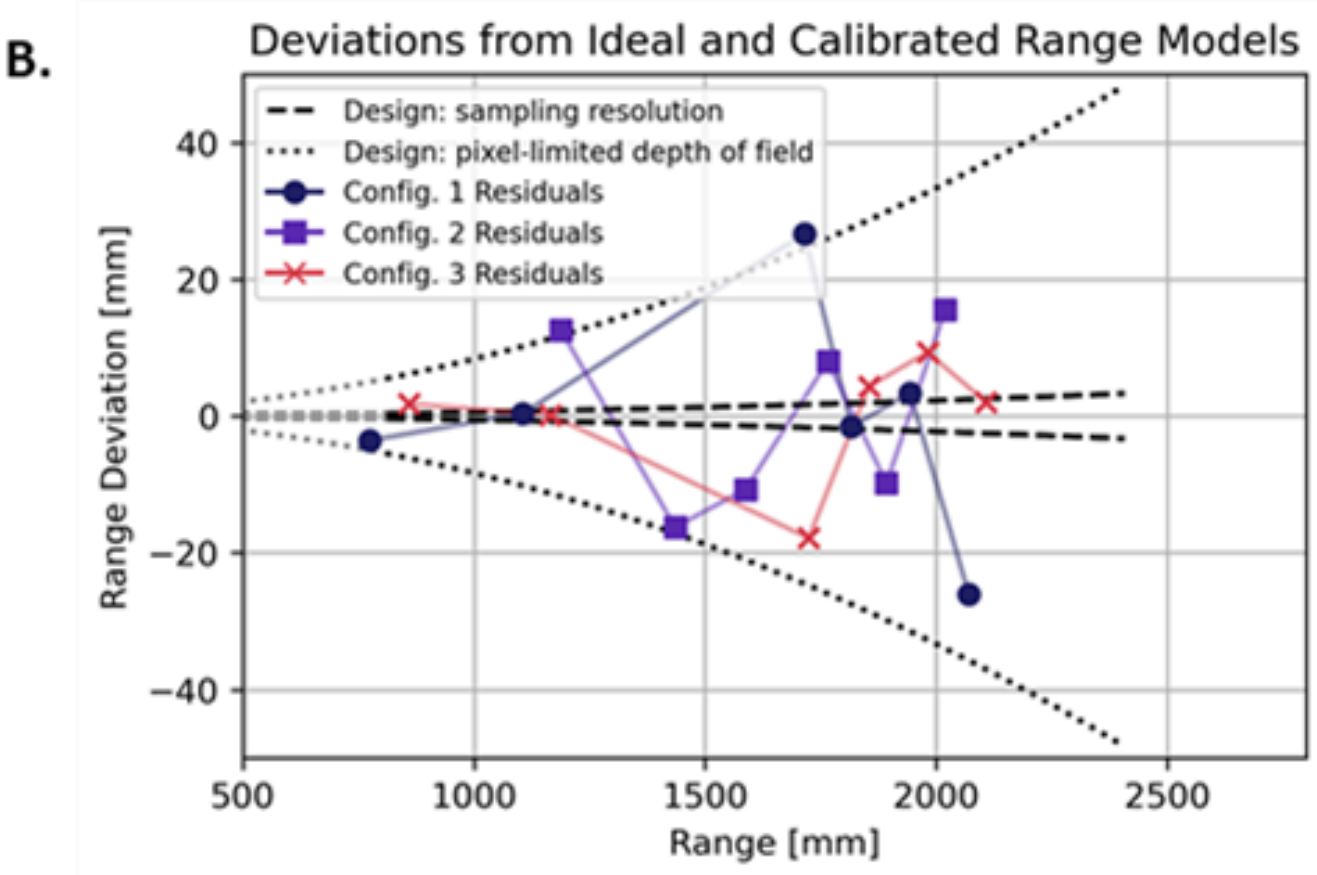

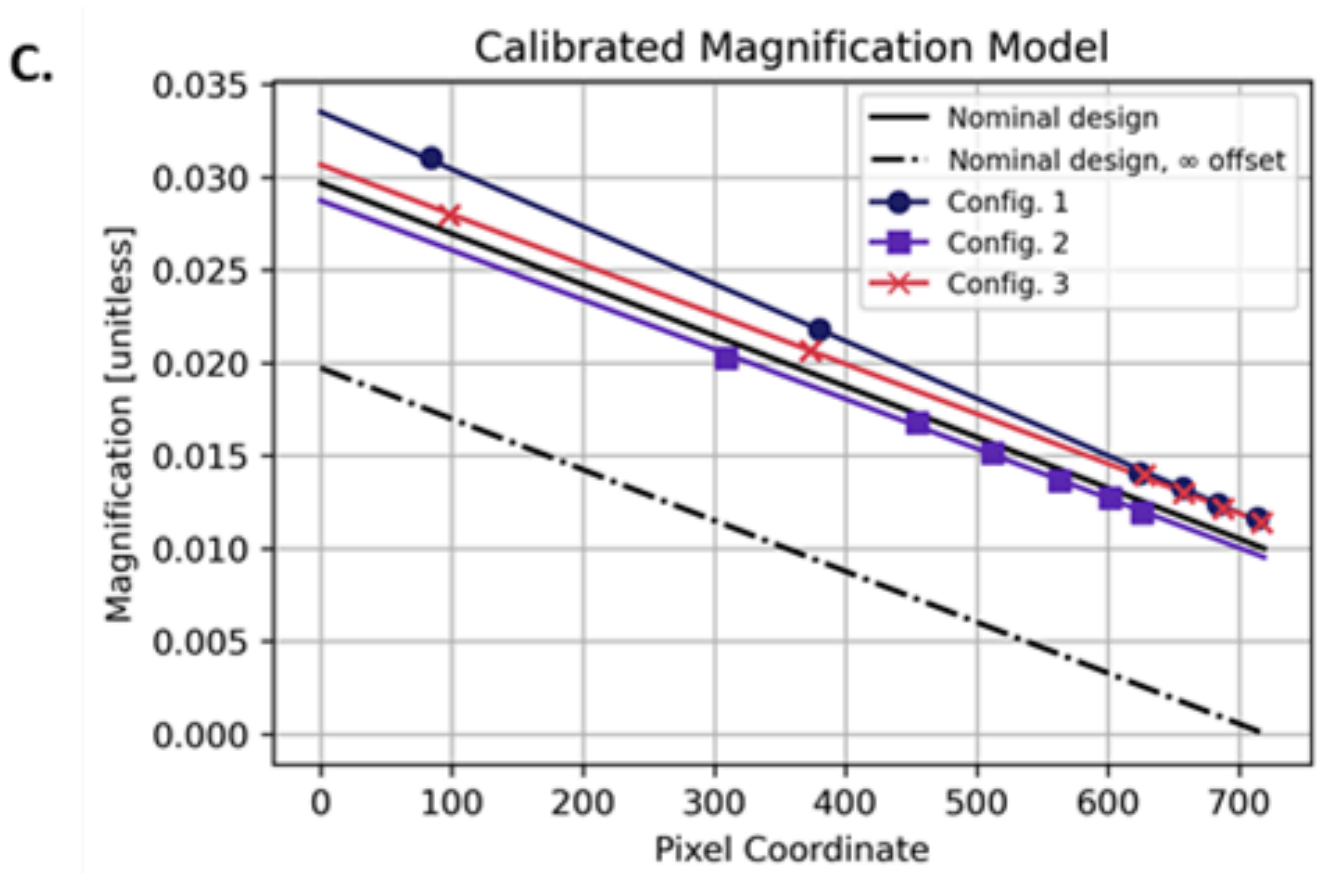

Fig. 4: Calibration Results (A) Pixel-to-Range maps showing measurement points for and fitted models for as-built configurations relative to nominal design model. (B) Residual fit errors for the models relative to system design limits for geometrical ranging uncertainty. (C) Pixel-to-Magnification map is an alternate method for assessing the as-built configurations relative to the nominal designs.

## 2.3. Point Cloud Generation

With the pixel-to-range mapping of Eq. (2) established, each sensor pixel corresponds to a calibrated object-plane depth. Converting the raw event stream into a 3D point cloud necessitates three sequential steps: noise suppression, centroid extraction, and coordinate assignment.

Raw event tensors are first passed through a background activity filter (BAF) implemented in SENPI [33, 36], which applies a spatiotemporal erosion operation to suppress isolated noise events that lack spatial or temporal neighbors. The filtered event tensor is then accumulated into a 2D projection where each column corresponds to a sensor pixel index and each row aggregates events over the scan interval for that lateral position.

The depth coordinate of the laser return at each column is recovered by centroid estimation along the range axis. Because event-based statistics approach a zero-mean Gaussian distribution for sufficient event counts [38], the centroid can be estimated as the column-wise mean of the event distribution, with the standard deviation serving as a per-column confidence measure. The resulting centroid map is then spatially smoothed to enforce surface continuity across adjacent scan positions, suppressing isolated outliers while preserving the illuminated profile. The centroid pixel index is then converted to a range value using **Eq. (2)**, and the scan position is determined from the calibrated offset from the stage. Event tensors are segmented into temporal windows and processed in parallel to reduce reconstruction time. Point clouds are visualized using CloudCompare [37].

Two independent sources of uncertainty govern the final ranging accuracy. The first is geometric uncertainty, which comprises the projected depth of field and the object-plane pixel binning. The depth-of-field contribution is given by **Eq. (3)**, and the discrete sampling contribution by:

$$\Delta R = \frac{R^2}{kk'} \Delta R' \tag{4}$$

with $\Delta R'$ being the sampling resolution on the image plane. For pixelated arrays $\Delta R'$ is one pixel, and k’ can be scaled by the pixel width, as in **Table 1**. For systems operating well beyond the front focal length, the depth-of-field term dominates, and the total geometric uncertainty is conservatively set as the maximum of the two contributions at each range. eSCHORTY is configured with a near 90° object plane tilt that makes the depth-of-field term the limiting factor across the full working range compared to the projected pixel size.

Detection uncertainty is the other source limiting accuracy. It arises from the finite laser beam divergence that causes the illuminated spot to subtend multiple pixel columns in event space, which creates a blooming effect that broadens the apparent centroid distribution and biases the range estimate. As characterized in **Sec. 3.2**, driving the laser modulation into breakdown progressively narrows the effective beam profile in event space by suppressing the low-intensity returns below the contrast threshold, which subsequently reduces the detection footprint to the peak of the profile. This modulation-breakdown trade-off provides a practical mechanism for reducing detection uncertainty at the cost of increased event rate, and its influence on ranging precision is quantified in the results.

To improve initial point cloud quality, we match the point generation rate to the laser modulation rate for temporal aggregation during each modulation period and apply event thresholding to remove background events. With this method, the maximum point cloud generation rate is the product of the laser modulation rate (e.g., 70 Hz for our experiments) and the number of scan angle samples (e.g., number of EVK4 rows, 1280).

## 3. Results

### *3.1 Experimental eSCHORTY Platform*

The eSCHORTY prototype, shown in **Fig. 5**, is constructed entirely from off-the-shelf optical and electronic components with a custom 3D-printed lens mount. An Apinex 650 nm CW line laser (LN-20, 20° fan beam truncated to 14°) provides the active illumination source, modulated by a UNI-T UTG9628 arbitrary waveform generator at selectable frequencies to drive continuous event generation. The Nikon Nikkor 24mm F/2 AI-s lens is mounted in a custom 3D-printed, light-tight housing mated directly to the Prophesee EVK4 event sensor, isolating the focal plane from stray ambient light. For scanning, the assembled unit is mounted on a Thorlabs DDS220 direct-drive linear stage, which translates the system across the scene at a fixed, encoder-tracked velocity to provide the along-track dimension of the point cloud.

**Table 2** summarizes eSCHORTY's performance when used as a pushbroom LiDAR where the single scan plane is swept across a scene by translating the entire system. The 1280 x 720 EVK4 sensor was oriented to provide 720 discrete range points within the working distance for each of the 1280 independent asynchronous scan angles across the camera's 14° view angle. The modulated 1 mrad x 14° illumination beam was aligned to fill the projected scan plane. The method for point cloud generation, as shown in **Fig. 3**, provides a maximum surface point generation rate based on the choice of laser modulation, which is generated by an estimated 89,600 points per second along a projected laser line at 70 hz (frequency used in data collect) and over 1.28M points at 1000+hz (high event-rate frequency from **Sec. 3.2**). However, conversion from events to points will vary based on smoothing parameters used in **Sec. 2.2**.

Table 2: eSCHORTY Performance as a Single Scan Plane “Pushbroom” LiDAR

| PERFORMANCE | eSCHORTY Values |
|---|---|
| Working Distance | 0.8 m to 2.5 m |
| Scan Angle | 14 deg Line |
| Depth Sampling Resolution | 0.9 mm at 0.8 m, 2.8 mm at 2.5 m (approx 0.1% m per m of Depth) |
| Along Track Angular Resolution | approx. 1 mrad |
| Across Track Angular Resolution | approx. 200 urad |
| Across Track Sampling | 1280 points |

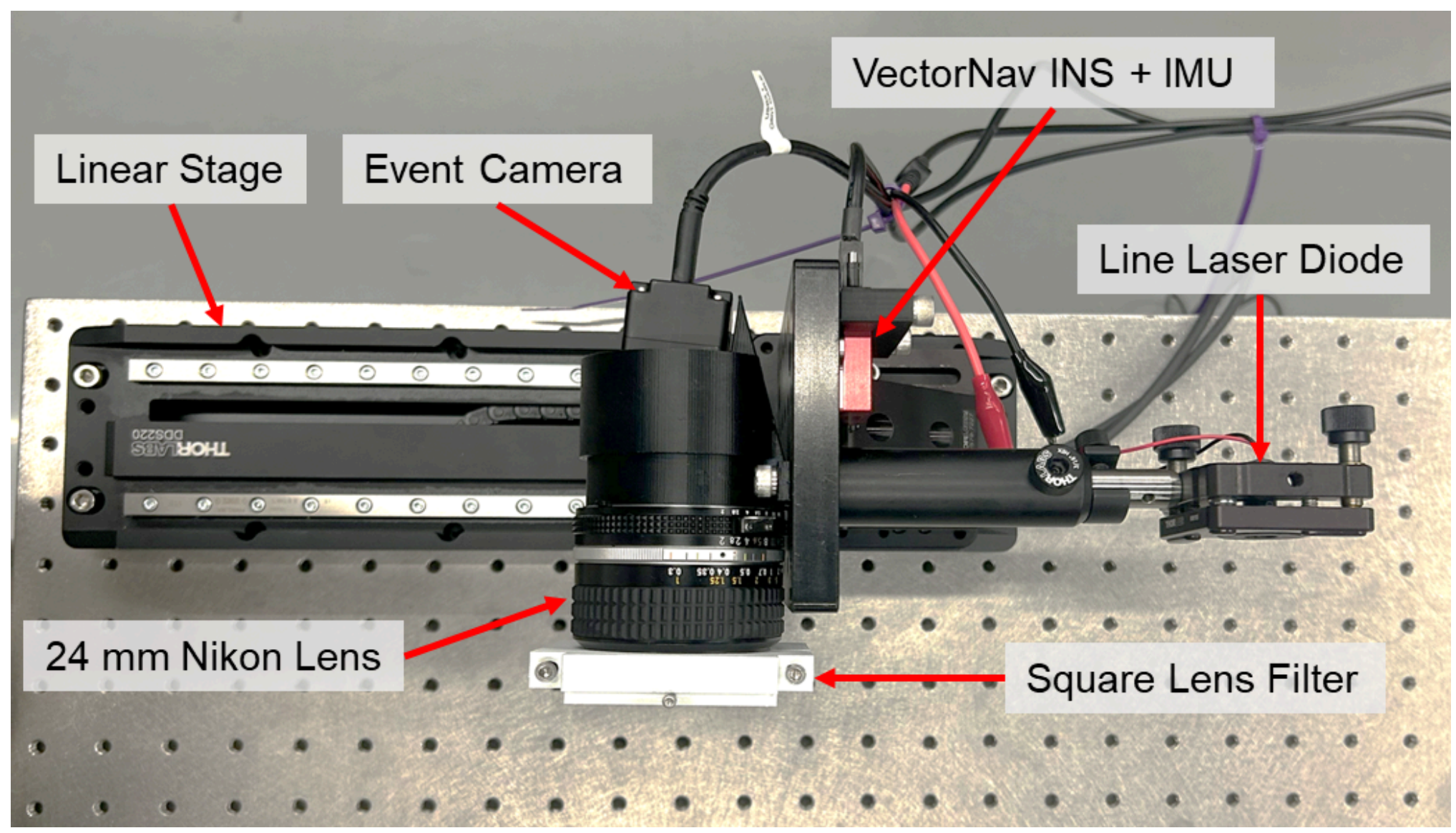

Fig. 5: eSCHORTY benchtop prototype system.

*3.2 Influence of Laser Modulation on Event Space Encoding*

The number of events eSCHORTY generates per scan, as well as the spatial profile underlying those events, depends jointly on the laser modulation frequency, the laser amplitude, and the EBS contrast threshold. Understanding this relationship is essential for setting the system to balance feature detectability against spatial precision across different operating environments. To characterize this trade space, we fix a flat white paper target at a known range and vary the modulation frequency from 50 Hz to 20,000 Hz at default EBS threshold settings to record both the total event count and the spatial laser profile at each frequency. The results are shown in **Fig. 6**.

The event response separates into three distinct regimes, as shown in **Fig. 6A**. At low modulation frequencies (e.g, 50–500 Hz), the full laser profile remains above the contrast threshold, and event count increases linearly with modulation rate, as each additional modulation cycle contributes a proportional number of detections. As modulation increases into the intermediate regime (e.g., 500–1000 Hz), the laser begins to break down. Here, the profile edges fall toward the threshold while the peak remains detectable, producing a saturating nonlinear increase in event count as the effective detection area narrows. At high frequencies (e.g., break down regime) (e.g., above ~1000 Hz), breakdown dominates and the laser intensity outside the central peak drops below the contrast threshold entirely, causing the event count to decrease while the remaining events are concentrated tightly along the peak of the spatial profile, as shown in **Fig. 6B**.

This three-regime behavior reflects a fundamental trade-off between feature detection and spatial localization in our active EBS configuration. The intermediate regime maximizes event count and correspondingly the feature detectability in low-contrast or high-background environments, while the break down regime modulation narrows the effective laser footprint in event space and improves ranging precision by concentrating detections at the intensity peak. As shown in **Fig. 6C**, the FWHM of the detected laser line decreases monotonically through the breakdown regime, directly reducing the detection uncertainty described in **Sec. 2.3**. Here, we observe a reduction in laser FWHM from ~20 rows to 2 rows as the laser approaches breakdown, creating an order of magnitude tighter accuracy on measured features using eSCHORTY. In practice, modulation frequency serves as the primary parameter to tune for this trade-off due to its direct impact on laser modulation and breakdown, with laser amplitude and EBS threshold providing secondary controls. This mechanism provides an additional degree of freedom for adapting eSCHORTY to environments ranging from high-background outdoor scenes to precision indoor profilometry.

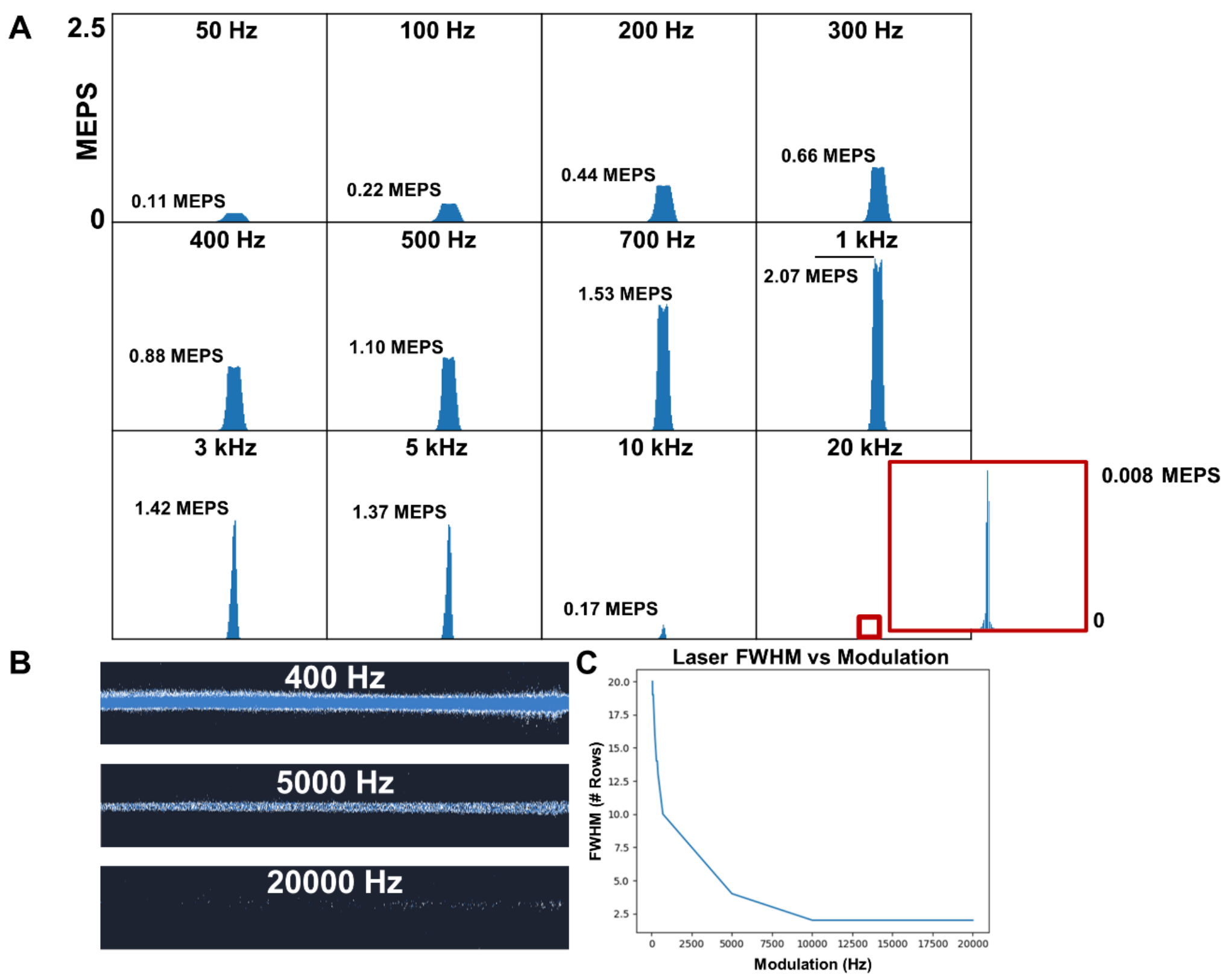

Fig. 6. (A) Influence of modulation rate on eSCHORTY measured events. (B) Example laser profile at variable modulation. (C) Laser line FWHM versus modulation rate.

### *3.3 Influence of Object Reflectance in Event Space Encoding*

Reflectance variation across a scene introduces artifacts in triangulation-based ranging due to asymmetric intensity shifts along tilted imaging geometries, displacing the measured centroid from the true surface position [40]. To characterize this effect in eSCHORTY, we interrogate contrast targets spanning the full reflectance range available described by their percent contrast from black and monitor the centroid position in event space at fixed range, as shown in **Fig. 7**.

Consistent with prior observations in intensity-based triangulation [40], eSCHORTY exhibits a centroid displacement that scales with the magnitude of the reflectance step, shown in **Fig. 7C**. However, because the EBS encodes brightness on a logarithmic scale, the displacement follows a logarithmic rather than linear relationship with reflectance, as shown in **Fig. 7F**. Across the contrast range tested, this logarithmic compression reduces the centroid displacement relative to the linear case, confirming that event-based detection is intrinsically less susceptible to this artifact than intensity-based centroiding. Here, we observe a maximum centroid shift of 3.75 pixels at the maximum contrast (e.g., 100%), whereas a linear relationship would scale this shift beyond 7.5 pixels (approximated by linearizing the first two points in **Fig. 7F**). While the measured displacement is minimized for EBS systems, the effect remains measurable and should be accounted for in high-accuracy ranging applications.

Beyond the centroid displacement, the microsecond temporal resolution of the EBS captures multiple event samples per scan point, providing direct access to per-point event statistics. Radiometrically, the laser return intensity follows an inverse-square law with range, while reflectance variations scale the return linearly. These dependencies produce distinguishable signatures when converted into event-space with range-induced drift follows $[-2\log(R) + \log(\sigma^2)]$ where $\sigma$ is the Gaussian beam divergence, while reflectance-induced drift follows $[-\log(R(x,y))]$ where $R(x,y)$ is the local surface reflectance. Considering that these two contributions drift at different rates suggests a path toward jointly solving for range and reflectance from the event statistics alone. As such, this result motivates this observation as an avenue for future work, particularly in multi-wavelength configurations that could enable event-based 3D reflectometry.

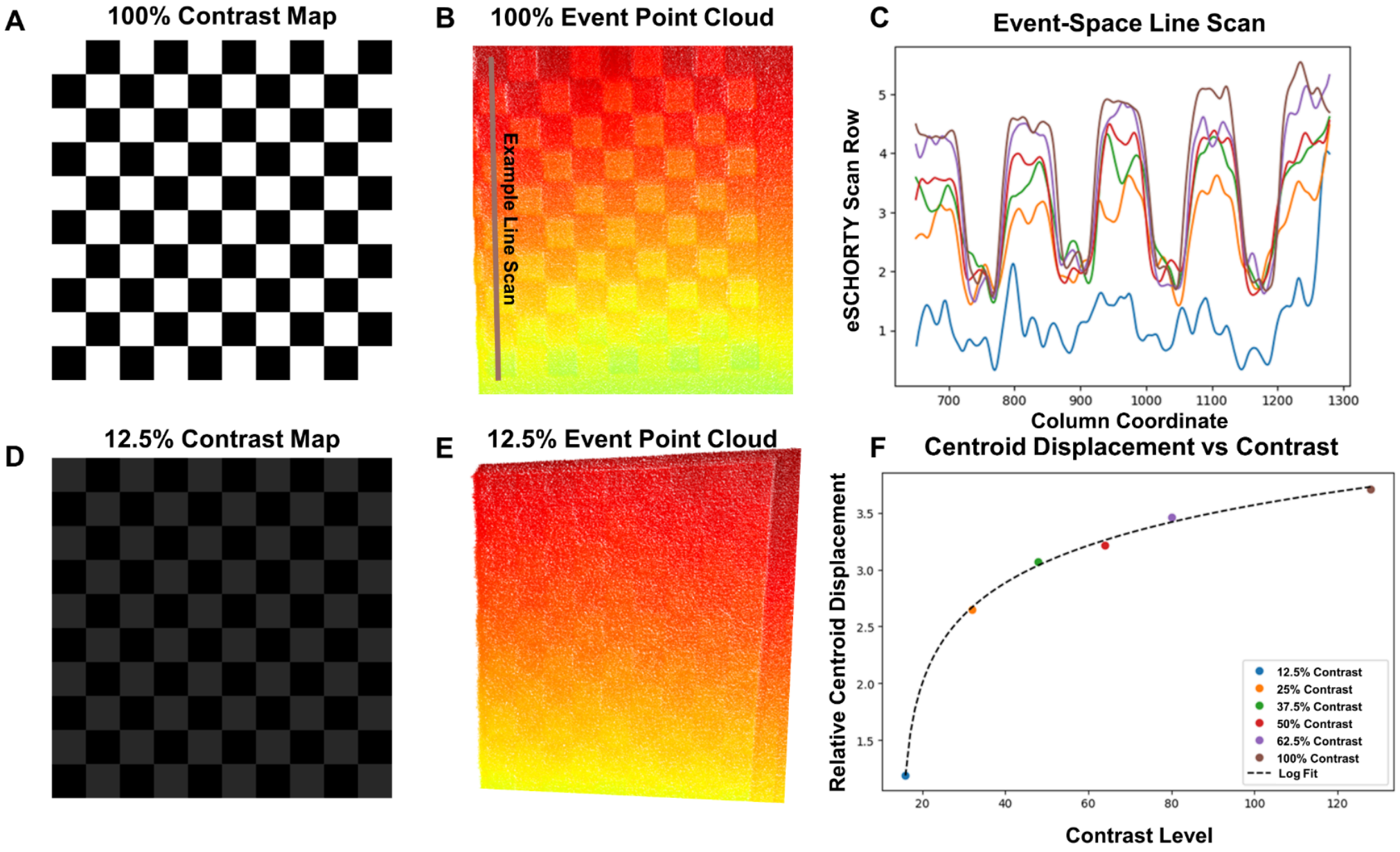

Fig. 7. (A) Example of a maximum contrast target in intensity-space and (B) event-space used for calibration. (C) The event-space displacement of the beam centroid as a function of different contrast levels. (D) Minimum contrast target in intensity- and (E) event-space. (F) Centroid statistical and fit as a function of contrast level. Legend in (F) also applies to (C).

### 3.4 Demonstration on Natural Scenes

To assess eSCHORTY's performance on uncontrolled targets, we scan two natural scenes: a collage of overlapping leaves arranged across multiple depth planes, and facial profiles of three co-authors at approximately one meter range. We display representative point clouds in **Fig. 8**.

The leaf collage, shown in **Fig. 8A–B**, spans a depth extent of approximately 210 mm across 3 overlapping layers. eSCHORTY resolves individual features separations down to 2.2 mm at the working distance of 1.32 m. A geometric consequence of the Scheimpflug architecture is that distinct depth planes map to distinct pixel rows on the tilted sensor, allowing for simultaneous capture of multiple layers without introducing range ambiguity. The only depth information lost is by direct occlusion, where foreground surfaces block the laser return from background features, which is a limitation shared by all single-viewpoint active systems.

Facial scans of three co-authors, shown in **Fig. 8C–E**, demonstrate eSCHORTY's ability to recover human-scale surface geometry, including nose bridge profile, and ear geometry at a range of 2.43 m. At this range, the across-track point spacing is 0.631 m and the along-track spacing is 0.355 m at the 70 Hz demonstrated scan rate. The EBC parameters for both experiments were adjusted such that events arising from the background were suppressed and eSCHORTY scanned at a rate of 300 mm/s.

These results confirm that eSCHORTY produces spatially continuous point clouds on natural scenes. The angular projection of the Scheimpflug geometry accommodates targets spanning a wide range of physical scales (e.g., from centimeter-scale leaf features to facial geometry) by adjusting working distance or lens focal length, as discussed in the calibration framework of **Sec. 2.2**.

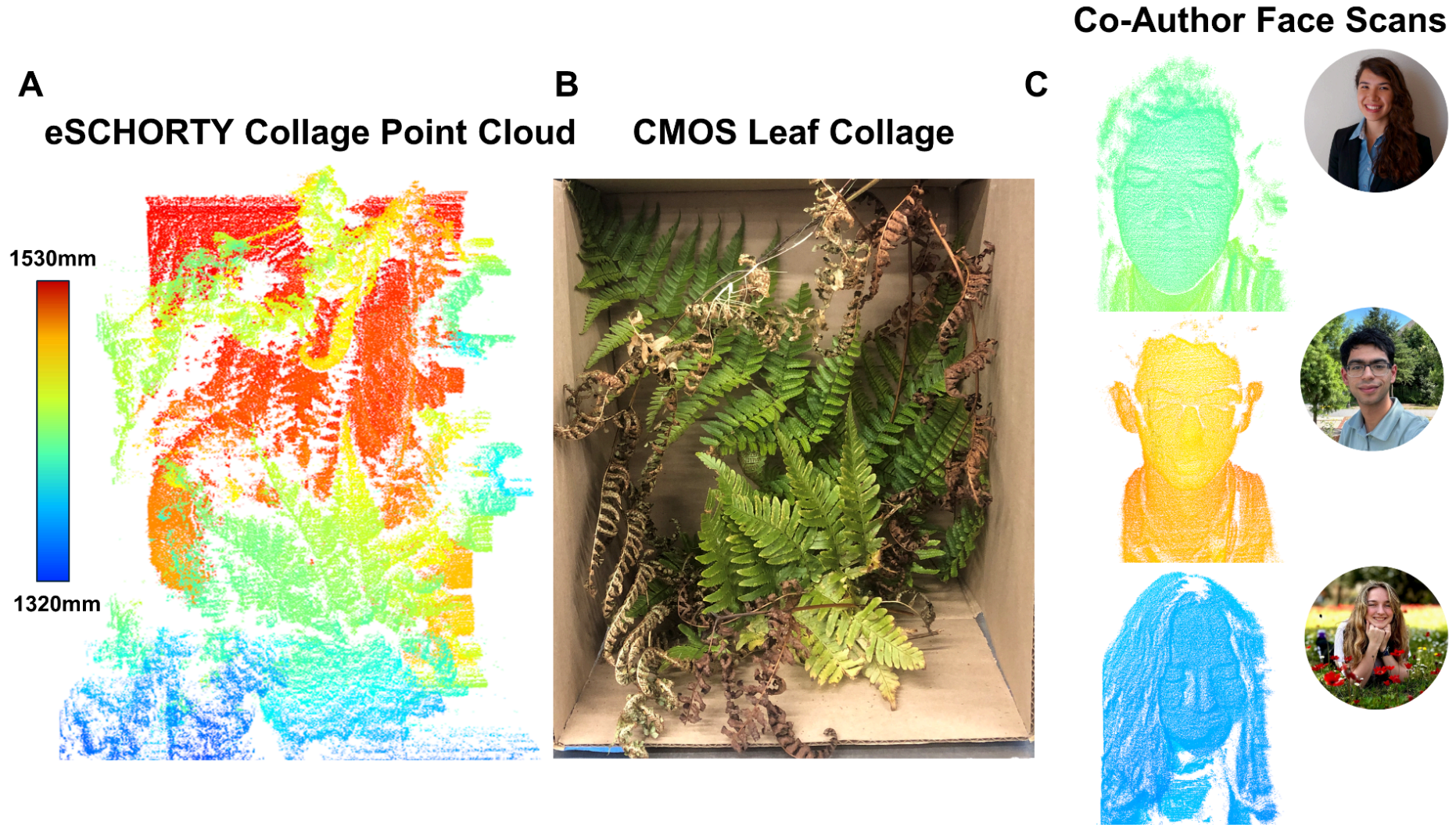

Fig. 8. (A) eSCHORTY point cloud of a leaf collage of overlapping natural objects. (B) CMOS reference image. (C) eSCHORTY scans of co-author faces.

## 4. Discussion

eSCHORTY demonstrates that integrating an event-based sensor into an active Scheimpflug ranging geometry provides a pushbroom LiDAR framework unconstrained by traditional framerate limits and with an observed contrast-driven mechanism for rejecting or accepting features. In contrast, prior Scheimpflug LiDAR systems [25,26,28] operate at frame rates typically below a few hundred Hz, limiting point cloud density at practical scan speeds. By encoding depth on a tilted sensor and reporting only contrast changes from a modulated laser line, eSCHORTY achieves dense, spatially continuous point clouds within the geometric depth-of-field bounds, while intrinsically suppressing static and out-of-focus returns. Within the 0.8–2.5 m working range (**Table 2**), this architecture delivers depth sampling on the order of 0.1% of range without time-gating or custom optics.

A key innovation is the study of laser modulation to tune the trade-off between feature detectability and spatial localization. The modulation study (**Sec. 3.2**) shows three regimes as frequency increases: linear growth of event count at low frequencies, saturation and onset of breakdown at intermediate frequencies, and a breakdown-dominant regime where the detectable footprint collapses to the beam peak. In the breakdown-dominant regime, the event-space FWHM of the laser line shrinks from ~20 to ~2 rows (**Fig. 6C**), tightening centroid localization by roughly an order of magnitude at the cost of event density. This behavior allows users to select between higher robustness (more events, broader footprint) and higher precision (fewer events, narrower footprint) with a single control parameter.

The reflectance experiment quantifies a second important trade-off. As in intensity-based triangulation [39], reflectance steps along the tilted geometry bias the centroid, but in eSCHORTY this bias scales logarithmically with contrast, yielding a maximum measured shift of 3.75 pixels over the tested range (**Fig. 7C,F**), roughly half of what a linear model would predict. This effect is caused by the logarithmic response of the event sensor, which reduces (but does not eliminate) reflectance-induced ranging error. For applications requiring sub-pixel accuracy across strongly varying materials, this residual bias is a clear limitation and will require either per-material calibration or algorithmic compensation. At the same time, the distinct analytical forms of range-induced and reflectance-induced drift [40] in event space, reinforced with the multiple events available per scan point to reduce noise to motivate a future joint inversion of range and reflectance, particularly in multi-wavelength configurations.

Throughput in the current prototype is practically limited by a low modulation (70 Hz), USB bandwidth, and the mechanical scan stage, resulting in demonstrated event rates from ~0.01 MEPS to ~1 MEPS and dense point clouds on natural scenes (**Sec. 3.4**). While simple estimates based on the EVK4 refractory period (e.g., estimated maximum event rate on the order of 100khz) and 1280 scan angles suggest a theoretical sensor-limited point generation ceiling approaching 128 MEPS, breakdown physics, background activity filtering, and centroid smoothing reduce this upper bound in practice.

Future iterations of this work will explore these practical tradeoffs to provide greater insights into how eSCHORTY may be optimized to maximize extraction of physical quantities and balance data rates with density. Multi-wavelength configurations will be explored and alternative waveforms and modulations may be studied to determine optimal event density given the log-encoding of EBCs. Additionally, we recognize that the Scheimpflug principle is limited by geometric bounds, which leads to an undesirable degradation in resolution as a function of range. Future iterations will explore how alternative centroid or algorithmic approaches may aid in breaking resolution limits to promote eSCHORTY as a high precision, long range tool. Finally, because the Scheimpflug condition and calibration procedure of **Sec. 2.2** are agnostic to range, the same architecture can scale, with appropriate changes to focal length and tilt, from centimeter-scale profilometry to kilometer-scale sensing as demonstrated in [29], promoting future applications using eSCHORTY as a reconfigurable event-based Scheimpflug LiDAR platform rather than a fixed-range instrument.

## Funding.

This project was supported by GTRI's Independent Research and Development (IRAD) and the GTRI Research Internship Program (GRIP) programs.

## Acknowledgment.

We additionally would like to thank Amanda Villegas, Ian Winski, and Claudia Vitale, and Yassine Fouchal for supporting the earliest efforts of this research as GTRI student interns. We also thank the GRIP team of Rossy Dang, Dillan Synan, Mikhail Velez, and Varun Vudathu for their inspiring research and for generating our first Scheimpflug lidar point cloud. We are grateful for our collaborations with Megan Birch, Denly Lindeman, and Katie Twitchell.

## Disclosures.

On behalf of all authors, the corresponding author states that there is no conflict of interest.

## Data Availability.

Data underlying the results presented in this paper are not publicly available at this time but may be obtained from the author upon reasonable request.

## Supplementary.

See Supplement 1 for supporting content

## References

[1] Y. Li and J. Ibanez-Guzman, "Lidar for Autonomous Driving: The Principles, Challenges, and Trends for Automotive Lidar and Perception Systems," *IEEE Signal Process. Mag.*, vol. 37, no. 4, pp. 50–61, July 2020, doi: 10.1109/MSP.2020.2973615.

[2] F. Bonin-Font, A. Ortiz, and G. Oliver, "Visual Navigation for Mobile Robots: A Survey," *J. Intell. Robot. Syst.*, vol. 53, no. 3, pp. 263–296, Nov. 2008, doi: 10.1007/s10846-008-9235-4.

[3] H. Lozano Bravo, E. Lo, H. Moyes, D. Rissolo, S. Montgomery, and F. Kuester, "A Methodology for Cave Floor Basemap Synthesis from Point Cloud Data: a Case Study of Slam-Based LIDAR at Las Cuevas, Belize," *ISPRS Ann. Photogramm. Remote Sens. Spat. Inf. Sci.*, vol. X-M-1–2023, pp. 179–186, June 2023, doi: 10.5194/isprs-annals-X-M-1-2023-179-2023.

[4] Institute of Geography, Pavol Jozef Šafárik University in Košice *et al.*, “Large-scale and high-resolution 3-D cave mapping by terrestrial laser scanning: a case study of the Domica Cave, Slovakia,” *Int. J. Speleol.*, vol. 44, no. 3, pp. 277–291, Sept. 2015, doi: 10.5038/1827-806X.44.3.6.
[5] D. Selvaratnam and D. Bazazian, “3D Reconstruction in Robotics: A Comprehensive Review,” *Comput. Graph.*, vol. 130, p. 104256, Aug. 2025, doi: 10.1016/j.cag.2025.104256.
[6] C. Poullis and S. You, “3D Reconstruction of Urban Areas,” in *2011 International Conference on 3D Imaging, Modeling, Processing, Visualization and Transmission*, Hangzhou, TBD, China: IEEE, May 2011, pp. 33–40. doi: 10.1109/3DIMPVT.2011.14.
[7] A. Christodoulides *et al.*, “Survey on 3D Reconstruction Techniques: Large-Scale Urban City Reconstruction and Requirements,” *IEEE Trans. Vis. Comput. Graph.*, vol. 31, no. 10, pp. 9343–9367, Oct. 2025, doi: 10.1109/TVCG.2025.3540669.
[8] R. Wang, J. Peethambaran, and D. Chen, “LiDAR Point Clouds to 3-D Urban Models: A Review,” *IEEE J. Sel. Top. Appl. Earth Obs. Remote Sens.*, vol. 11, no. 2, pp. 606–627, Feb. 2018, doi: 10.1109/JSTARS.2017.2781132.
[9] L. Zhou, G. Wu, Y. Zuo, X. Chen, and H. Hu, “A Comprehensive Review of Vision-Based 3D Reconstruction Methods,” *Sensors*, vol. 24, no. 7, p. 2314, Apr. 2024, doi: 10.3390/s24072314.
[10] J. Butime, I. Gutierrez, L.G. Corzo, and C.F. Espronceda, “3D Reconstruction Methods, A Survey,” in *Proceedings of the First International Conference on Computer Vision Theory and Applications*, Setúbal, Portugal: SciTePress - Science and and Technology Publications, 2006, pp. 457–463. doi: 10.5220/0001369704570463.
[11] M. F. M. Costa, “Surface inspection by an optical triangulation method,” *Opt. Eng.*, vol. 35, no. 9, p. 2743, Sept. 1996, doi: 10.1117/1.600840.
[12] D. Scharstein and R. Szeliski, “High-accuracy stereo depth maps using structured light,” in *2003 IEEE Computer Society Conference on Computer Vision and Pattern Recognition, 2003. Proceedings.*, Madison, WI, USA: IEEE Comput. Soc, 2003, p. I-195-I–202. doi: 10.1109/CVPR.2003.1211354.
[13] M. Hansard, S. Lee, O. Choi, and R. Horaud, *Time-of-Flight Cameras: Principles, Methods and Applications*. in SpringerBriefs in Computer Science. London: Springer, 2013. doi: 10.1007/978-1-4471-4658-2.
[14] S. Yang and G. Zhang, “A review of interferometry for geometric measurement,” *Meas. Sci. Technol.*, vol. 29, no. 10, p. 102001, Oct. 2018, doi: 10.1088/1361-6501/aad732.
[15] D. Huang *et al.*, “Optical Coherence Tomography,” *Science*, vol. 254, no. 5035, pp. 1178–1181, Nov. 1991, doi: 10.1126/science.1957169.
[16] S. R. P. Pavani *et al.*, “Three-dimensional, single-molecule fluorescence imaging beyond the diffraction limit by using a double-helix point spread function,” *Proc. Natl. Acad. Sci.*, vol. 106, no. 9, pp. 2995–2999, Mar. 2009, doi: 10.1073/pnas.0900245106.
[17] C. Wang, G. Ballard, R. Plemmons, and S. Prasad, “Joint 3D localization and classification of space debris using a multispectral rotating point spread function,” *Appl. Opt.*, vol. 58, no. 31, p. 8598, Nov. 2019, doi: 10.1364/AO.58.008598.
[18] C. Liu *et al.*, “Enhancing the resolution of three-dimensional migration images based on space-variant point spread function deconvolution,” *Geophys. Prospect.*, vol. 72, no. 5, pp. 1672–1697, June 2024, doi: 10.1111/1365-2478.13477.
[19] Kelly, D. P., Healy, J. J., Hennelly, B. M. and Sheridan, J. T. (2011) *Quantifying the 2.5D imaging performance of digital holographic systems.* Journal of the European Optical Society, 6 (11034). ISSN 1990-2573

[20] F. Aguet, D. Van De Ville and M. Unser, "Model-Based 2.5-D Deconvolution for Extended Depth of Field in Brightfield Microscopy," in *IEEE Transactions on Image Processing*, vol. 17, no. 7, pp. 1144-1153, July 2008, doi: 10.1109/TIP.2008.924393.
[21] J. Greene et al. “Pupil engineering for extended depth-of-field imaging in a fluorescence miniscope”, *Neurophotonics* vol.10 iss. 4 (2023)
[22] B. Javidi *et al.*, “Multidimensional Optical Sensing and Imaging System (MOSIS): From Macroscales to Microscales,” *Proc. IEEE*, vol. 105, no. 5, pp. 850–875, May 2017, doi: 10.1109/JPROC.2017.2654318.
[23] J. Wu *et al.*, “An integrated imaging sensor for aberration-corrected 3D photography,” *Nature*, vol. 612, no. 7938, pp. 62–71, Dec. 2022, doi: 10.1038/s41586-022-05306-8.
[24] Scheimpflug, Theodore, “Improved method and apparatus for the systematic alteration or distortion of plane pictures and images by means of lenses and mirrors for photography and for other purposes,” 1196, Jan. 16, 1904 [Online]. Available: http://trenholm.org/hmmerk/TSBP.pdf
[25] M. Brydegaard, E. Malmqvist, S. Jansson, G. Zhao, J. Larsson, and S. Török, “The Scheimpflug lidar method,” in *Lidar Remote Sensing for Environmental Monitoring 2017*, U. N. Singh, Ed., San Diego, United States: SPIE, Aug. 2017, p. 17. doi: 10.1117/12.2272939.
[26] F. Gao, H. Lin, K. Chen, X. Chen, and S. He, “Light-sheet based two-dimensional Scheimpflug lidar system for profile measurements,” *Opt. Express*, vol. 26, no. 21, p. 27179, Oct. 2018, doi: 10.1364/OE.26.027179.
[27] L. Wang *et al.*, “Research on application of Scheimpflug condition in rail profile measurement,” in *Seventh Symposium on Novel Photoelectronic Detection Technology and Applications*, J. Chu, Q. Yu, H. Jiang, and J. Su, Eds., Kunming, China: SPIE, Mar. 2021, p. 354. doi: 10.1117/12.2587466.
[28] K. Chen, F. Gao, X. Chen, Q. Huang, and S. He, “Overwater light-sheet Scheimpflug lidar system for an underwater three-dimensional profile bathymetry,” *Appl. Opt.*, vol. 58, no. 27, p. 7643, Sept. 2019, doi: 10.1364/AO.58.007643.
[29] N. D. Meraz *et al.*, “Scheimpflug cameras for range-resolved observations of the atmospheric effects on laser propagation,” in *Laser Radar Technology and Applications XXX*, M. D. Turner, G. W. Kamerman, and L. A. Magruder, Eds., Orlando, United States: SPIE, May 2025, p. 11. doi: 10.1117/12.3054806.
[30] P. Lichtsteiner, C. Posch, and T. Delbruck, “A 128 x 128 120 dB 15 microsecond Latency Asynchronous Temporal Contrast Vision Sensor,” *IEEE J. Solid-State Circuits*, vol. 43, no. 2, pp. 566–576, 2008, doi: 10.1109/JSSC.2007.914337.
[31] Z. Qu, Z. Zou, V. Boominathan, P. Chakravarthula, and A. Pediredla, “Event fields: Capturing light fields at high speed, resolution, and dynamic range,” Dec. 09, 2024, *arXiv*: arXiv:2412.06191. doi: 10.48550/arXiv.2412.06191.
[32] G. Gallego *et al.*, “Event-Based Vision: A Survey,” *IEEE Trans. Pattern Anal. Mach. Intell.*, vol. 44, no. 1, pp. 154–180, Jan. 2022, doi: 10.1109/TPAMI.2020.3008413.
[33] J. L. Greene, A. Kar, I. Galindo, E. Quiles, E. Chen, and M. Anderson, “A PyTorch-enabled tool for synthetic event camera data generation and algorithm development,” in *Synthetic Data for Artificial Intelligence and Machine Learning: Tools, Techniques, and Applications III,* Orlando, United States: SPIE, May 2025, p. 13. doi: 10.1117/12.3053238.
[34] J. E. Greivenkamp, *Field Guide to Geometrical Optics*. SPIE, 2004. doi: 10.1117/3.547461.
[35] I. Baldwin and P. Newman, “Road vehicle localization with 2D push-broom LIDAR and 3D priors,” in *2012 IEEE International Conference on Robotics and*

*Automation*, St Paul, MN, USA: IEEE, May 2012, pp. 2611–2617. doi: 10.1109/ICRA.2012.6224996.
[36] A. Lakshmi, A. Chakraborty, and C. S. Thakur, “Neuromorphic vision: From sensors to event-based algorithms,” *WIREs Data Min. Knowl. Discov.*, vol. 9, no. 4, p. e1310, July 2019, doi: 10.1002/widm.1310.
[37] *CloudCompare*. [GPL Software]. Retrieved from: http://www.cloudcompare.org
[38] R. Cao, D. Galor, A. Kohli, J. L. Yates, and L. Waller, “Noise2image: noise-enabled static scene recovery for event cameras” in *Optica*, vol. 12, iss 1, p. 45 (2025).
[39] B. Curless and M. Levoy, “Better optical triangulation through spacetime analysis” in International Conference for Computer Vision. (1995).
[40] M. Wan, S. Wang, H. Zhao, H. Jia, and L. Yu, “Robust and accurate sub-pixel extraction method of laser stripes in complex circumstances” in *Applied Optics*, vol. 60, *iss* 36, p. 11196 (2021).